\documentclass[aps,prl,showpacs,twocolumn,letterpaper]{revtex4}
\usepackage{graphicx}
\usepackage{subfigure}
\usepackage{amsmath}
\usepackage{bm}
\begin{document}

\title{Dynamical Structure, Bonding, and Thermodynamics of the Superionic 
Sublattice in $\alpha$-AgI}

\author{Brandon C. Wood}
\author{Nicola Marzari}
\affiliation{Department of Materials Science and Engineering, Massachusetts 
Institute of Technology, Cambridge, MA 02139}

\begin{abstract}
Using extensive first-principles molecular dynamics calculations, we 
characterize the superionic phase transition and the lattice and electronic 
structures of the archetypal Type-I superionic conductor $\alpha$-AgI. We find 
that superionicity is signalled by a phase transition of the silver ions alone.
In the superionic phase, the first silver shell surrounding an iodine displays 
a distinct dynamical structure that would escape a time-averaged 
characterization, and we capture this structure in a set of ordering rules. The
electronic structure of the system demonstrates a unique chemical signature of 
the weakest-bound silver in the first shell, which in turn is most likely to 
diffuse. Upon melting, the silver diffusion decreases, pointing to an unusual 
entropic contribution to the stability of the superionic phase.
\end{abstract}


\maketitle

Key advances in energy research have prompted a surge of interest in superionic
materials, as a crucial enabling technology for a variety of nanotechnological 
devices, including sensors, switches, batteries, and fuel cells. Of the 
superionics, AgI and related silver halides and sulfides have attracted 
particular attention because of the unusually high levels of ionic 
conductivity they exhibit, and as such are finding increased and varied 
technological implementation~\cite{terabe05,lee04,minami96}. At normal pressure,
AgI enters its superionic $\alpha$ phase above $T_{c}=420$~K, at which 
temperature a phase transition to a body-centered cubic structure is 
accompanied by an increase in the silver conductivity of nearly three orders of 
magnitude, to a value of 1.31~$\Omega^{-1}\thinspace\text{cm}^{-1}$~\cite{
kvist68}. Previous molecular dynamics studies using classical pair potentials 
have successfully reproduced experimental characteristics of the $\alpha$ and 
$\beta$ phases, as well as the $\alpha\rightarrow\beta$ transition~\cite{
vashishta78_parrinello83,tallon88,shimojo91,osullivan91,zimmer00,reviews}, but 
these are unable to describe the electronic structure in a dynamic environment, 
or to capture the phenomenology of the melting transition. In this regard, 
first-principles simulations provide unique and unbiased predictive power.

We perform Car-Parrinello molecular dynamics simulations in the canonical 
\textit{NVT} ensemble at temperatures ranging from 200~K to 1250~K for a total
of 800 ps.  All simulations were performed with a 54-atom unit cell
and $a_0=5.174$~\AA, except the Wannier function calculations, which were 
performed in a 32-atom unit cell~\footnote{Simulations were performed in a 
plane-wave basis set using the PBE-GGA XC functional, a $4d^{10}5s^{1}$ 
ultrasoft silver and $5s^{2}5p^{5}$ norm-conserving iodine pseudopotential, 
cutoffs of 22~Ry and 176~Ry for the wavefunctions and charge density, and a 
timestep of 20~au.}.

\begin{figure}
	\centering
	\includegraphics[angle=270]{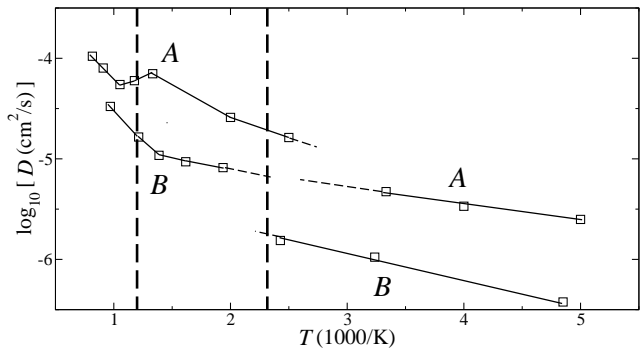}
	\caption{Arrhenius plot of the diffusion coefficient for the fully 
	mobile system (\textit{A}) and with the iodine sublattice fixed 
	(\textit{B}). The dashed vertical lines indicate the experimental 
	$T_c$ (420~K) and $T_m$ (830~K) at standard pressure.}
	\label{fig:arrhenius}
\end{figure}

First, we find evidence of a phase transition of the silver ions near the 
experimental $T_c$ that is independent of the conformation and dynamics of the 
iodine sublattice and signals the transition into the superionic $\alpha$ 
phase. The silvers exhibit a sharp decrease in their diffusion behavior upon 
cooling below $T_c$, although cubic boundary conditions forbid the iodine 
structural transition to the hexagonal wurtzite $\beta$ phase. Results from a 
series of simulations in which we immobilized the iodines in a fixed bcc 
configuration provide a secondary, stronger indicator of the independence of 
the silver transition from any iodine dynamics. Fig.~\ref{fig:arrhenius} 
displays the associated silver ion diffusion coefficients $D_{Ag}$ for the 
fixed-iodine case and for a fully mobile lattice. Diffusion coefficients are 
derived from the mean-square displacement (MSD) via the Einstein relation, 
$D=\lim_{t\rightarrow\infty}\frac{1}{6t}\left\langle MSD(t)\right\rangle $. In
both cases, the slope of an Arrhenius plot of $D_{Ag}$ shows a characteristic 
discontinuity near the experimental transition temperature. Unexpectedly, this 
discontinuity is even more pronounced for the fixed-iodine case. Immobilizing 
the iodine sublattice does lead to an overall decrease in $D_{Ag}$, suggesting 
local lattice fluctuations beneficial for silver mobility are frozen out, but 
the system retains its superionic behavior. 

Also of note in Fig.~\ref{fig:arrhenius} is the unusual decrease of the silver 
ion diffusion coefficient upon melting at about 850~K, a tendency that has also
been observed experimentally~\cite{kvist68,araki99} but is not captured by 
classical potentials. The thermodynamic implication gives valuable insight 
into the stability of the superionic phase. The decreased mobility of the 
silver ions upon melting signals a decrease in the silver entropic 
contribution, meaning the iodines crystallize into the superionic state below 
$T_m$ to increase the entropy of the silvers and make them more ``liquid''. 
Accordingly, the superionic phase acts as an intermediate between a pure solid 
and a pure liquid, with a high-entropy liquid sublattice flowing through a 
low-energy solid matrix. An entropically driven stabilization of the $\alpha$
phase is also consistent with experimental results~\cite{biermann60}, which 
determine the entropy difference between superionic and liquid phases to be 
relatively tiny.

\begin{figure}
	\centering
	\includegraphics[angle=270]{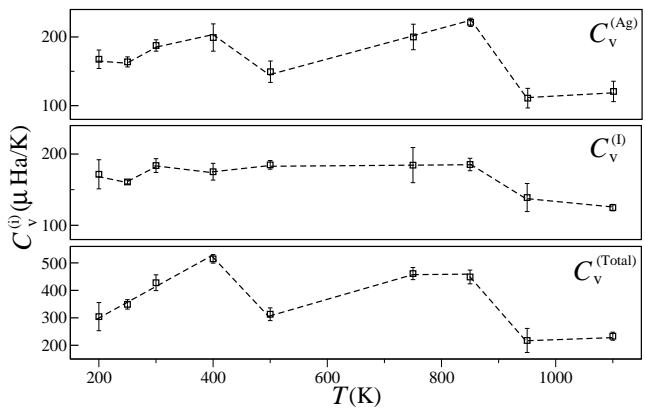}
	\caption{Specific heat capacities for the silver sublattice, the iodine
	sublattice, and the total system.}
	\label{fig:cv_ion}
\end{figure}

Although it is not possible to separate the total energy of the system into 
respective sublattice contributions, we can get a quantitative picture of the 
energetic fluctuations associated with each sublattice by instead integrating 
the forces on the ions, which are already decoupled. Since in the canonical 
ensemble, fluctuations $\sigma$ in the total energy are related to the 
specific heat capacity $C_{V}$ as $\sigma^2=\text{k}_BT^2C_V$, we can derive a 
heat capacity-like quantity $C_V^{(i)}$ that is sublattice resolved using the 
energetic fluctuations derived from the ionic forces. A plot of this quantity 
for the two sublattices, along with the total specific heat capacity for the 
entire system, is shown in Fig.~\ref{fig:cv_ion}. Over the full temperature 
range of the simulations, we observe the tail end of a divergence in the total 
heat capacity at 400~K and 850~K. These temperatures correspond closely to the 
superionic transition temperature $T_c$ and the melting temperature $T_m$, 
respectively.  Examination of the associated ion-resolved $C_V^{(i)}$ curves 
reveals a similar divergence for the silvers at $T_c$ but no detectable trend 
for the iodines. As in Fig.~\ref{fig:arrhenius}, we can thus link the 
superionic transition to the silver sublattice only. 

\begin{figure}
	\centering
	\includegraphics{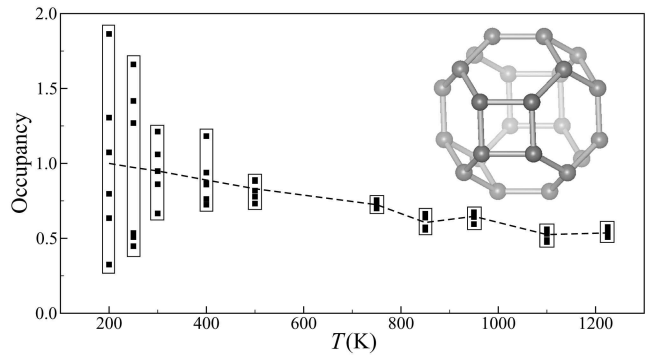}
	\caption{Average silver occupancies of the six inequivalent tetrahedral
	bcc interstitial sublattices of Ref.~\cite{szabo86}, normalized
	against the case for which all ions occupy random tetrahedral sites. 
	The inset shows the network of tetrahedral sites surrounding a single 
	iodine, and the dashed line indicates the overall fraction of silvers 
	in tetrahedral sites.}
	\label{fig:tetr_occ}
\end{figure}

We note that an order-disorder transition of the silver ions at $T_c$ has been 
reported previously~\cite{madden92,seok97_98}. Our results confirm separation 
of the occupied tetrahedral sites into the six inequivalent sublattices of 
Ref.~\cite{szabo86}, the occupancies of which are plotted as a function of 
temperature in Fig.~\ref{fig:tetr_occ}. The simulations reveal an ordering 
tendency for the silvers below $T_c$, characterized by a splitting into higher- 
and lower-occupancy sublattices. The independent ordering tendency of the silver 
sublattice upon cooling is observable despite the inhibition of the 
$\alpha\rightarrow\beta$ transition, in agreement with Ref.~\cite{
madden92}. However, at these low temperatures, even more extensive statistical
sampling is necessary to allow for comparison of the ordered phase with those 
proposed in the cited studies.

\begin{figure}
	\centering
	\includegraphics[]{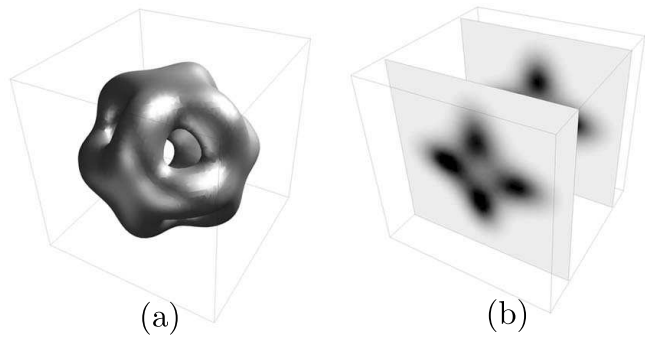}
	\caption{(a) The isosurface of silver trajectories at 750 K, enclosing 
	a volume associated with the first silver shell surrounding an iodine. 
	(b) A slice plot of (a), revealing higher density (darker) silver 
	occupancy regions.}
	\label{fig:ag_contours}
\end{figure}

We have also investigated the most frequented pathways for the silver ions by 
tracking their positions with respect to the iodines and averaging the resulting
trajectories over the course of the simulations. Fig.~\ref{fig:ag_contours} shows
isosurface and slice plots for the silver ion density in the first shell 
surrounding an iodine. Our results confirm that the highest-density regions lie
near the tetrahedral sites, with silver ion density smeared toward the 
octahedral sites. This agrees well with the experimental results of 
Ref.~\cite{wuensch77}.

\begin{figure}
	\centering
	\includegraphics[angle=270]{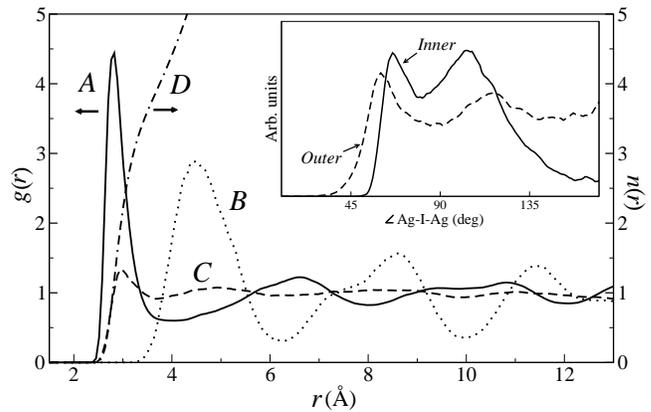}
	\caption{Radial pair distribution functions for (\textit{A}) Ag--I, 
	(\textit{B}) I--I, and (\textit{C}) Ag--Ag, along with the integrated 
	Ag--I $n(r)$ curve (\textit{D}), at 750~K. Inset shows the angular 
	distributions of the four silver ions in the first shell surrounding an
	iodine, measured with respect to the remaining silvers in the shell and
	resolved according to the three innermost silvers and the fourth outer
	silver.}
	\label{fig:ag_dist}
\end{figure}

From the ionic trajectories, we can identify a set of rules governing the 
instantaneous distribution of silvers. The silver-silver pair distribution 
function $g_{Ag-Ag}(r)$ (Fig.~\ref{fig:ag_dist}) illustrates a zero probability 
of finding silver ions closer together than $R_{Ag-Ag}=2.4$~\AA, which precludes 
simultaneous occupation of nearest-neighbor tetrahedral interstitial sites. 
Integration of $g_{Ag-I}(r)$ over the first peak reveals that on average, each 
iodine has four nearest neighbors in the first shell, consistent with geometric 
expectations for tetrahedral interstitial site occupancy. In addition, 
$g_{Ag-I}(r)$ indicates the highest probability distance for silvers surrounding 
an iodine is at $R_{Ag-I}=2.6$~\AA, which corresponds to the radial distance to 
a tetrahedral interstitial site. However, a time-resolved analysis suggests a 
more detailed decomposition of the first peak of $g_{Ag-I}(r)$. We find that 
most commonly, three of the four nearest-neighbor silver ions simultaneously 
occupy a shell that corresponds to the tetrahedral site distance (the average 
value varies from 2.7 to 3.0 depending on temperature, with lower temperatures 
favoring higher values). The fourth silver is seen to transition regularly 
between this shell and an equivalent shell for a neighboring iodine such that
on average, it fills the transition zone between the two, as defined by 
$2.8 \leq R_{Ag-I} \lesssim 4.2$~\AA. The transitioning rate is temperature 
dependent and disappears rapidly for $T<T_c$. The fourth silver also possesses 
an angular distribution distinct from its three inner counterparts, as indicated 
in the inset of Fig.~\ref{fig:ag_dist}. The Ag--I--Ag angles for any of the 
closest three silvers reveal preferences at $65^{\circ}$ and $105^{\circ}$, and 
higher angles are surprisingly uncommon. However, the angles introduced by the 
inclusion of the fourth nearest-neighbor shell are comparatively diffuse and 
have significant probabilities towards larger values. This suggests that 
whereas the closest three silvers are clustered and correlated in their 
positions, the fourth silver is relatively unconstrained in its angular 
configuration and is affected only marginally by the orientations of the 
remaining three. The rare permanent transitions of this unconstrained 
fourth silver between shells represent the driving factor for mass diffusion, 
an observation that would escape experimental investigations.

The above analysis allows for the definition of a set of ordering rules that 
govern the instantaneous distribution of silver ions in the first shell 
surrounding an iodine: (1) four silver ions populate the first shell; (2) no 
two silver ions occupy neighboring tetrahedral interstitial sites; (3) on 
average, three silver ions surround an iodine at a radius of 
$R_{Ag-I}=2.6$~\AA, the tetrahedral interstitial distance; (4) a fourth 
silver transitions between that shell and a second shell at 
$R_{Ag-I}\gtrsim4.2$~\AA, associated with a neighboring iodine, and the 
transition rate between the two is temperature dependent and disappears below 
$T_c$; (5) the angular positions of the three inner silvers are correlated, 
with preferred Ag--I--Ag angles of $65^{\circ}$ and $105^{\circ}$, whereas the 
fourth (outer) silver is relatively unconstrained; and (6) silver ions tend to 
organize into high- and low-occupancy sublattices at temperatures below $T_c$.

\begin{figure}
	\centering
	\includegraphics[]{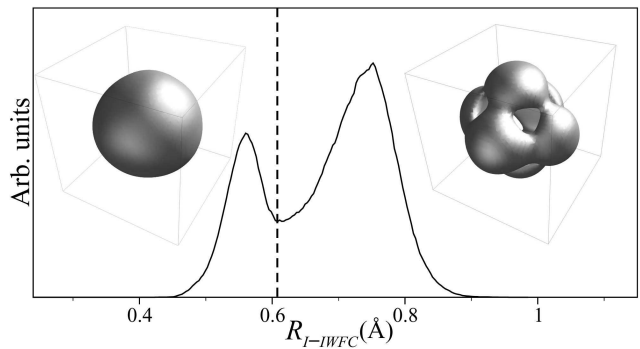}
	\caption{Histogram of distances between iodine WFCs and iodine nuclei. 
	The insets are contours of the iodine WFC distribution about their 
	nuclei for WFCs on either side of the broken line.}
	\label{fig:iiwfc_contour}
\end{figure}

We examined the maximally localized Wannier functions (MLWFs) to obtain a local 
picture of bonding in a dynamic environment~\cite{marzari97}. Plotting the 
time-averaged radial distribution of the four iodine Wannier function centers 
(WFCs) about their parent ion reveals an unexpected bimodal separation into 
short-distance, highly localized WFCs and longer-distance, partially delocalized 
WFCs (Fig.~\ref{fig:iiwfc_contour}). Further isolating Wannier functions 
associated with each of the two peaks and plotting the time- and 
statistical-averaged contours for the WFCs around the iodines yields the 
isosurfaces shown in the insets of Fig.~\ref{fig:iiwfc_contour}. The 
long-distance Wannier function centers (LWFCs) tend to align along the cubic 
axes toward the octahedral interstitial sites. Their orientations relate to the 
observed smearing of the silver occupancy from the electrostatically preferred 
tetrahedral sites towards the octahedral face centers and suggest that these 
orbitals correspond to directional interactions between silvers and iodines. On 
the other hand, the short-distance iodine Wannier function centers (SWFCs) 
exhibit a random angular distribution, as would be expected in a strictly 
Coulombic picture. Comparing the areas of the two peaks in 
Fig.~\ref{fig:iiwfc_contour} reveals that of the four WFCs surrounding an 
iodine, 30\% on average can be classified as SWFCs, a figure which agrees with 
the likelihood for a first-shell silver to be found in the mobile transition 
zone. This value also correlates well with the the experimentally determined 
likelihood of a silver to be located outside the tetrahedral interstitial 
site~\cite{wuensch77,nield93}, a quantity confirmed by our findings. 

\begin{figure}
	\centering
	\includegraphics[angle=270]{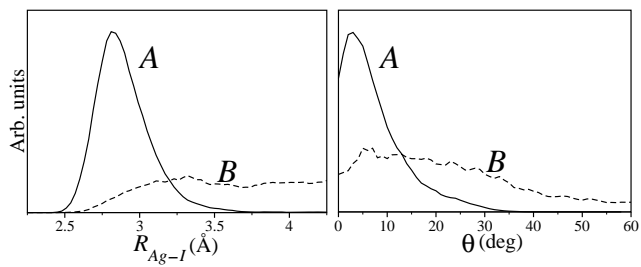}
	\caption{Histograms of interionic bond distance and bond angle for the 
	silvers closest to iodine WFCs, resolved into (\textit{A}) 
	long-distance and (\textit{B}) short-distance Wannier functions. 
	The bond angle $\theta$ is defined as the angle between 
	$\bm{R}_{Ag-I}$ and $\bm{R}_{IWFC-I}$.}
	\label{fig:iwfc}
\end{figure}

A more detailed picture of the correlation between delocalization extent of the
iodine MLWFs and the positions of nearby silver atoms is offered in 
Fig.~\ref{fig:iwfc}. There is clear evidence of a chemical interaction between 
LWFCs and silver atoms within a threshold radius of 
$R_{Ag-I}\lesssim3.25$~\AA\ and a solid bond angle of 
$\lvert\theta\rvert\lesssim20^{\circ}$. This positional constraint on the 
silvers indicates the bonding with LWFCs is directional and not merely 
Coulombic. As a quantitative measure of these more complex chemical 
interactions, we have also calculated the Born effective charges for the ionic 
sublattices, using the electric-enthalpy method of Refs.~\cite{umari02_souza02}. 
We obtain values of $Z^*=\pm 1.22$, larger than the integral values expected 
for a purely ionic crystal, further supporting the conclusions of our Wannier 
function analysis. Fig.~\ref{fig:iwfc} shows no such directional interaction 
for SWFCs, indicating that corresponding interactions with silvers can be 
attributed to weaker electrostatics. Moreover, examination of the radial 
distribution for first-shell silvers closest to SWFCs places them within the 
transition zone. 

We conclude that silvers contributing to $D_{Ag}$ are those which are not bound
to LWFCs, meaning they are minimally constrained both radially and angularly. 
Most commonly, the outermost of the four first-shell silvers fills these 
criteria, as a lack of stronger directional interactions with LWFCs increases 
the average Ag--I bond distance. This comparatively unconstrained, mobile fourth
silver fleetingly occupies the transition zone defined by 
$2.8 \leq R_{Ag-I} \lesssim 4.2$~\AA\ until it is captured by a neighboring 
iodine, leading to mass diffusion. At higher temperatures, thermal disordering 
breaks a greater number of bonds between silvers and LWFCs, promoting more
nearby silvers into the transition zone. As such, the overall fraction of 
occupied tetrahedral sites decreases (Fig.~\ref{fig:tetr_occ}) and diffusion 
is enhanced.

In conclusion, we have shown that the transition to the superionic $\alpha$
phase of AgI is signalled by an independent phase transition of the silver 
sublattice alone, characterized by a disordering of the silvers and a sharp 
increase in their diffusivity. Upon melting, the silver diffusion coefficient
decreases, pointing to an unusual entropic contribution to the stabilization of 
the superionic phase. We have 
also identified diffusion pathways for superionic silver ions above $T_c$, and 
a time-resolved analysis of ion trajectories has allowed us to define a set of 
ordering rules that govern the instantaneous distribution of silvers in the 
first shell surrounding an iodine. Finally, we have found that of the four 
first-shell silvers, the closest three are strongly correlated and restricted 
in their angular distribution, and that they are involved in anisotropic, 
directional bonding to an iodine. The fourth silver is bound only weakly and 
is relatively unconstrained, and we have isolated it as the dominant 
contributor to mass diffusion.

\begin{acknowledgments}
Funding for this work has been provided by the DOE CSGF fellowship and MURI 
Grant DAAD 19-03-1-0169. Calculations have been done with the Quantum-ESPRESSO
package~\cite{espresso} using computational facilities provided through NSF 
grant DMR-0414849. The authors also wish to thank Prof. Bernhard Wuensch 
for helpful discussions.
\end{acknowledgments}

\end{document}